\documentclass[conference]{IEEEtran}
\usepackage{url}
\usepackage{algorithmic}
\usepackage{graphicx}
\usepackage{textcomp}
\usepackage{xcolor}
\usepackage{listings}
\usepackage{mathtools}
\definecolor{codegray}{gray}{0.9}
\newcommand{\code}[1]{\colorbox{codegray}{\texttt{#1}}}
\usepackage{flushend}

\AtBeginDocument{%
  }

\begin{document}

%%
%% The "title" command has an optional parameter,
%% allowing the author to define a "short title" to be used in page headers.
\title{KModels: Unlocking AI for Business Applications}

\author{%
Roy Abitbol\textsuperscript{1,*}, 
Eyal Cohen\textsuperscript{1}, 
Muhammad Kanaan\textsuperscript{1}, 
Bhavna Agrawal\textsuperscript{2}, 
Yingjie Li\textsuperscript{2}, 
Anuradha Bhamidipaty\textsuperscript{2},
Erez Bilgory\textsuperscript{1}
\\[1ex]
\textsuperscript{1}IBM Research, Israel \quad
\textsuperscript{2}IBM Research, USA
% \\[1ex]
}

\maketitle

\renewcommand{\thefootnote}{\fnsymbol{footnote}}
\footnotetext[1]{Corresponding author: Roy Abitbol (roy.abitbol@il.ibm.com)}
\renewcommand{\thefootnote}{\arabic{footnote}}

\begin{abstract}
    As artificial intelligence (AI) continues to rapidly advance, there is a growing demand from clients and product managers to integrate AI capabilities into their existing business applications. However, a significant gap exists between the rapid progress in AI and the slow rate at which AI is being embedded into business environments. Deploying well-performing lab models into production settings, especially in on-premise environments, often entails specialized expertise and imposes a heavy burden of model management, creating significant barriers to implementing AI models in real-world applications.
    
    KModels is built on the shoulders of proven libraries and platforms (Kubeflow Pipelines, KServe) aiming to simplify AI adoption by supporting both the AI developers and the consumers. It allows model developers to focus their efforts solely on model development and to share the models as transportable units (Templates), abstracting away complex production deployment concerns. At the same time, KModels allows AI models consumers to eliminate the need for a dedicate data scientist, since the templates encapsulate most data science considerations while still providing business-oriented control.

    In this paper we discuss the architecture of KModels and the key decisions that helped shape it. We present KModels' main components and their function as well as its interfaces. Furthermore, we explain how KModels is highly suited for on-premise deployment but can also be used in cloud environments.
    
    The efficacy of KModels is demonstrated through the successful deployment of three different AI models within an existing Work Order Management system. These models are deployed in a client's data center and are trained successfully on local data, without any data scientist intervention. One of these models improved the accuracy of Failure Code specification for work orders from 46\% to 83\%, showcasing the substantial benefit of accessible and localized AI solutions.
\end{abstract}

\section{Introduction}
    
    The rapid advancements in artificial intelligence (AI) have generated significant interest from clients and product managers who seek to integrate AI capabilities into their business applications. Success stories of embedding AI into cloud business applications (Software-as-a-Service) are common \cite{lins2021artificial}, however, the data in cloud applications is controlled by the software vendors and the AI models are trained, deployed and maintained by them (This is especially true for consumer applications but in this paper we focus on business oriented application). The ability to access all clients data allows the vendors an ideal setup where they can perfect the data cleansing and preparation and eventually optimize the resulting models. Vendors may offer the best suited model for each subset of clients or even train a dedicated model per each client.
    
    On the other hand, many companies are running business applications on-premises, due to cost, privacy, security or regulatory reasons \cite{kemp2018legal}. In these settings the data is kept locally and cannot be shared outside. As a result the software vendors of such applications, do not have any access to client data, and are essentially prohibited from training and providing any AI models to their clients. In practice, many application vendors delegate the responsibility of training AI model to their clients by shipping AI models as raw code \cite{chui2018ai}. However, this approach places the burden of training the model on the clients and requires each client to employ AI engineers (or data scientists). This burden effectively increases the cost of ownership of any new AI model thus setting a high bar for the adoption of AI \cite{ransbotham2019winning,lee2020multi}. Companies lacking the financial resources and the suitable skills required for this task will not be able to apply and utilize new AI capabilities.
   
    One of the major challenges is the need to manage the entire life-cycle of the AI models. In order to successfully and continuously run AI models in production environments, it is necessary to employ a range of technical and operational services and skills to support the models deployment, connectivity, monitoring and many other aspects of their life-cycle. The technical effort involved in this is surprisingly high \cite{sculley2015hidden} and it has yielded an entirely new area of operational AI, known as DevOps for AI \cite{lwakatare2020devops} or MLOps \cite{alla2021mlops}. In a typical cloud application the developers manage the entire life cycle of the AI models. However, if the model is run on-premises of the client's data center (especially if the model is delivered as raw code), the client engineers are burdened with managing the entire life cycle of the model.

    \begin{figure*}
    \center\includegraphics[width=0.8\linewidth]{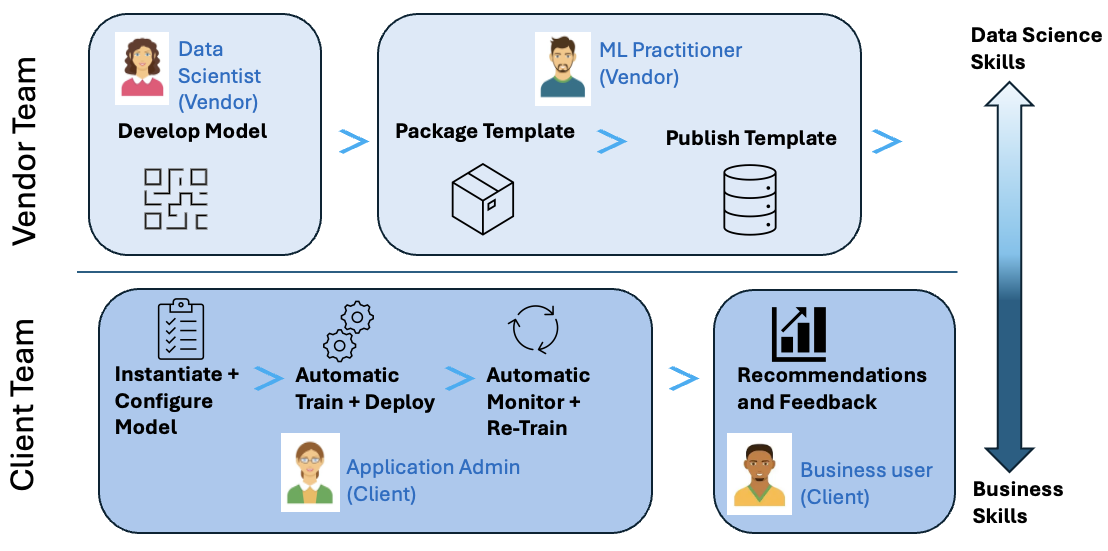}
    \caption{The entire life span of an AI model as envisioned by KModels. It starts with the development of the model at the lab, continues with its packaging and publishing as a template. Both these steps are done on the model vendor's end and require skilled personnel . Later, on the client's end the model is instantiated, trained and deployed, requiring only business configuration and no data science skills.}
    \label{fig:kmodels-flow}
    \end{figure*}

    To recap, the key challenges inhibiting widespread adoption of AI in business applications are:
    \begin{itemize}
        \item Adapting the AI models to perform well on local client environments, requires sharing the models code in order to train them on-premises.
        \item Training the models on clients premises, requires skilled data-scientists, increasing the cost of running AI models.
        \item Models that are running in local environments require constant monitoring and management, entailing a heavy burden for local operational staff.
    \end{itemize}
        
    These challenges present a barrier for incorporating AI into business applications and environments, leading to a relatively low success rate of AI projects and low adoption of new AI models. According to a research by Gartner Inc. \cite{costello2020gartner} "only 53\% of projects make it from AI prototypes to production" and at a recent McKinsey global survey \cite{panikkar2021mckinzey} only 36\% of respondents said that "ML algorithms had been deployed beyond the pilot stage".

    Reviewing the yearly \textit{Global AI Adoption Index} \cite{ibm2024morning} \cite{ibm2021morning} over the past 3 years, shows that enterprise IT professionals have persistently ranked the accessibility and simplicity of AI as the top driver of AI adoption in their organizations (``more accessible and easier to deploy''). Furthermore, when asked about the main barriers to successful AI adoption, they consistently point, year after year, towards the scarcity of ``AI skills, expertise or knowledge''. Other top barriers mentioned were complexity of the deployment and difficulty to integrate and scale. These studies are a strong testament to the validity and prevalence of the challenges at hand. It is evident that despite the decades of work on AI development there is still room for improvements in the operational aspects of AI, in order to allow broad acceptance of AI in business applications and environments.

    In this paper, we present KModels, a novel framework aimed at reducing the technical skills required for AI integration in business applications, without sacrificing quality or comprehensiveness. On one hand, KModels offers the application developers a mechanism to ship their models to clients local premises such that the models will be trained and deployed in a controlled and manageable way. On the other hand, KModels allows the application clients to adopt new AI models with minimal effort and without a skilled AI personnel.
    
    The key contributions made by this paper are:
    \begin{enumerate}
        \item Introducing KModels, a framework that simplifies the deployment and life-cycle management of AI models. KModels allows non-technical users to deploy and operate models in production environments while preserving the clients' data locally, on-premises.
        \item Discussing the architecture of KModels and analyzing the key decisions that helped shape it. We present KModels' main components and their function as well as their interfaces.
        \item Lastly, explaining how KModels is highly suited for on-premise local deployments and why KModels infrastructure can enable the widespread adoption of AI throughout various business applications.
    \end{enumerate}

\section{Related Work}
\label{related}
The landscape of AI frameworks, which is evolving both in the academia and in the commercial AI industry, is overwhelmed with numerous solutions for different use cases and needs. These frameworks can be broadly categorized into AI Libraries, AutoML platforms, and MLOps frameworks. While each category addresses specific challenges in the AI lifecycle, none provide the comprehensive, end-to-end solution for business users that KModels offers.

\begin{table*}[ht]
\centering
\caption{Comparison of KModels with other AI development and deployment frameworks}
\label{tab:comparison}
\resizebox{\textwidth}{!}{%
\begin{tabular}{lccccc}
\hline
\textbf{Feature} & \textbf{KModels} & \textbf{\begin{tabular}[c]{@{}c@{}}AI Libraries\\ (e.g., PyTorch, TensorFlow)\end{tabular}} & \textbf{\begin{tabular}[c]{@{}c@{}}AutoML\\ (e.g., Google AutoML)\end{tabular}} & \textbf{\begin{tabular}[c]{@{}c@{}}MLOps Frameworks\\ (e.g., Kubeflow)\end{tabular}} \\
\hline
Low-code model deployment & $\surd$ & $\times$ & $\surd$ & $\times$ \\
On-premises deployment & $\surd$ & $\surd$ & $\times$ (typically cloud-based) & $\surd$ \\
End-to-end lifecycle management & $\surd$ & $\times$ & Partial & $\surd$ \\
Business-centric configuration & $\surd$ & $\times$ & Partial & $\times$ \\
Template-based model sharing & $\surd$ & $\times$ & $\times$ & $\times$ \\
No data science skills required for use & $\surd$ & $\times$ & $\surd$ & $\times$ \\
Customizable for specific business needs & $\surd$ & $\surd$ & Limited & $\surd$ \\
Built-in scalability & $\surd$ & $\times$ & $\surd$ & $\surd$ \\
Local data privacy & $\surd$ & $\surd$ & $\times$ (typically cloud-based) & $\surd$ \\
\hline
\end{tabular}%
}

\vspace{0.5em}
\raggedright
\footnotesize
This comparison of key features highlights how KModels combines the strengths of various approaches while addressing critical gaps in existing solutions, particularly for on-premises business applications.
\end{table*}

\subsection{AI Libraries}
AI libraries (such as PyTorch \cite{paszke2019pytorch,ansel2024pytorch}, TensorFlow \cite{tensorflow2015-whitepaper}, Scikit-Learn \cite{scikit-learn}, etc.) provide AI model developers the building blocks to develop and deploy models through a high-level programming interface \cite{nguyen2019machine}. These libraries contribute various Machine Learning (ML) and AI algorithms and provide developers high level functions and classes \cite{paszke2019pytorch}, thus placing an abstraction on top of low level coding tasks. These libraries require advanced coding and data science skills and are used to orchestrate high level algorithmic pipelines. While some of these libraries do provide certain deployment support \cite{goldsborough2016tour}, it is usually very limited, and not suitable for operational methodology. In Contrast, KModels is primarily focused on orchestrating the life cycle of the AI models but it is not concerned at all with the development process of the model. KModels is agnostic to the underlying AI libraries used in each model.

\subsection{Auto ML}
Auto-ML or Auto-AI solutions (such as IBM's AI Studio \cite{autoai2024ibm}, Google's AutoML \cite{automl2024google} and others) automate the construction of ML pipelines \cite{he2021automl} \cite{xanthopoulos2020putting}. They are designed to reduce the demand for data scientists and enable domain experts to automatically build ML applications without requiring deep statistical or ML knowledge. They operate by allowing the user to merely make a minimal set of decisions and operations while they apply extensive internal logic for analyzing the data and applying the best AI algorithm with the most suitable set of parameters. Like KModels, these frameworks promote the proliferation of AI models in business environments. They provide the infrastructure and means for business users who are less skilled in data science to easily train models \cite{bisong2019google}. However, these frameworks lack in their support for a scalable distribution of the models. Furthermore, unlike KModels, these frameworks are not equipped to deal with the entire management of models and their lifecycle

\subsection{MLOps Frameworks}
MLOps frameworks or ML orchestration frameworks (such as Kubeflow, Airflow or MLflow) target the challenge of managing complex model training pipelines in production \cite{hewage2022machine} \cite{kreuzberger2023machine}. They provide a set of building blocks allowing an operations engineer to construct a flow starting from the data collection through the model training and ending in model serving. In fact, Kubeflow pipelines are used by KModels to orchestrate the training process for the models inside KModels. However, these models require an expert level knowledge in their operation and configuration. Setting up models and continuously monitoring them using these tools requires a significant ongoing effort from skilled personnel. KModels abstracts away the underlying complexity of these systems behind a simple, template-driven interface purpose-built for business users. It takes the ML pipelines tested in the lab via MLOps tools and seamlessly deploys them to production with minimal configuration. 

By building on the shoulders of proven libraries and platforms (Figure \ref{fig:kmodels-stack}), KModels offers a uniquely accessible and business-friendly solution for end-to-end AI deployment and lifecycle management. Its template-driven approach, automated model ops, and self-serve configuration put the power of AI directly in the hands of business users while leveraging best-of-breed MLOps tools behind the scenes. This empowers enterprises to overcome the traditional barriers to AI adoption and rapidly infuse AI across their business applications and processes.

As shown in Table 1, KModels uniquely combines features that are typically spread across different types of frameworks. It offers the ease of use associated with AutoML solutions, the customizability and on-premises capabilities of traditional AI libraries, and the lifecycle management of MLOps frameworks. Crucially, KModels adds business-centric configuration and template-based model sharing, making it particularly well-suited for widespread AI adoption in business applications without requiring extensive data science expertise.

\section{KModels}
KModels was developed from the grounds up aiming to simplify AI adoption by supporting both the AI developers and the consumers. KModels allows model developers to focus their efforts solely on model development and allows the developers to package the models as transportable units (Template, discussed in section \ref{templates}), abstracting away complex production deployment concerns. At the same time, KModels allows AI models consumers to eliminate the need for a dedicate data scientist, since the templates encapsulate most data science considerations while still providing business-oriented control.

\subsection{Assumptions and Requirements}

As we set out to develop the KModels solution we devised several key assumptions that clarify the scope of the framework and the use cases it addresses. These assumptions are:

\begin{itemize}
    \item \textit{Business need} – As discussed in the introduction, we posit that there is a growing need for a solution that will make AI accessible to many clients as well as business applications, especially in on-premise deployments. Existing AI deployment mechanisms fall short of providing a comprehensive solution to the challenge (discussed in section \ref{related}). On-premise AI models are either one-off models, which are highly customized to a specific need, or model-building code releases requiring deep data-science and programming knowledge.
    \item \textit{Feasibility and applicability} – There are certainly complex cases mandating that the AI models will be manually tuned by a data scientist. However, we argue that in a significant share of the worthy business use cases, it is feasible to train AI models, on a client's data, autonomously, without requiring manual tuning by a data scientist.
    \item \textit{Integration with user interface} - The proposed solution deals with most aspects of AI models management in a generic fashion, however the solution does not provide any UI support for the inferencing of the models. The business applications shall access a REST endpoint to run the models inferencing and they are responsible for reflecting the AI output in any way or triggering any action as a result of the AI output. In other words, integrating the AI output into the application is not part of the scope of the current solution.
\end{itemize}

\begin{figure}
\center\includegraphics[width=\linewidth]{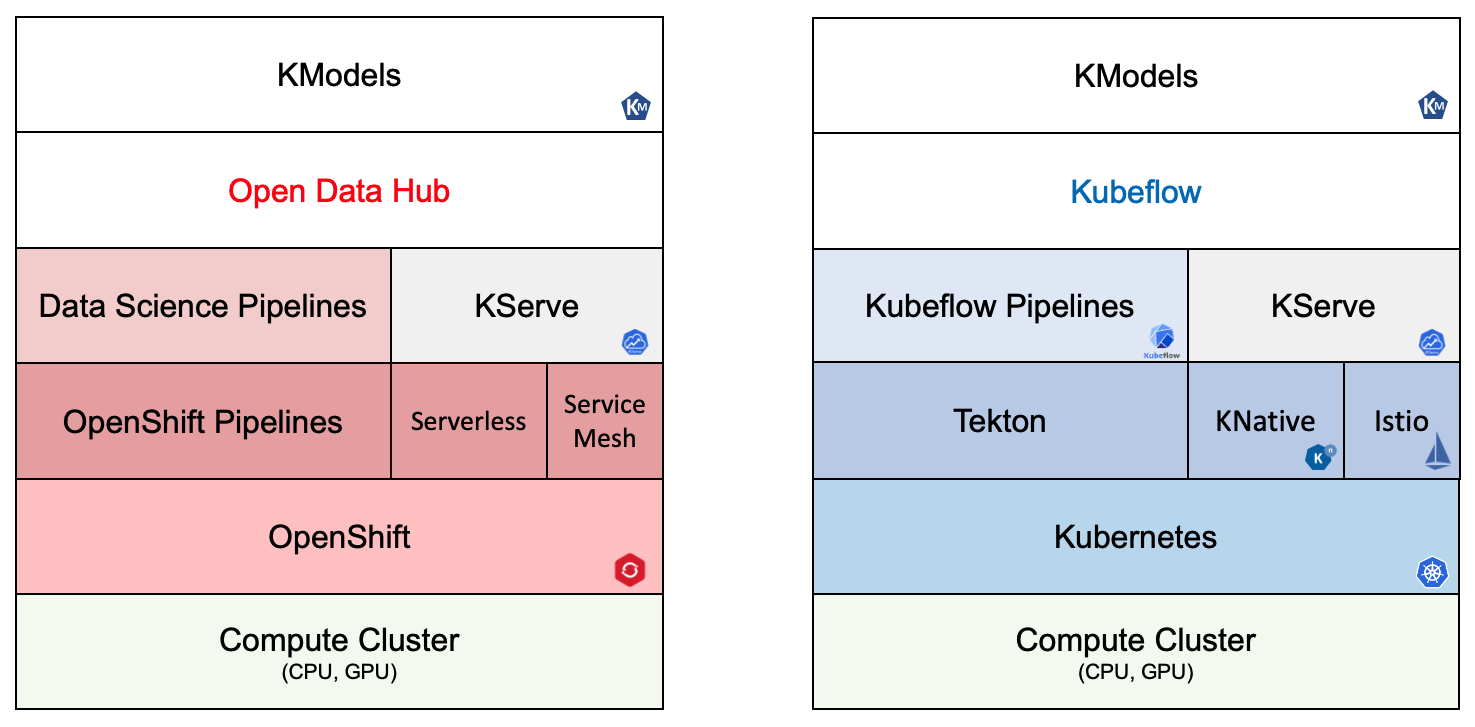}
\caption{Architecture diagram depicting KModels as the control layer on top of two different options of stacks: Kubernetes based or OpenShift based. The key components are KServe, and Kubeflow/Data-Science pipelines. The underlying infrastructure layers assure the robustness of KModels while the complexity of managing and configuring those layers is largely hidden by KModels' control layer, reducing the expertise level required from the target users.}
\label{fig:kmodels-stack}
\end{figure}

In addition to these assumptions we accumulated key requirements, over the course of defining the solutions, that have helped shape the solutions and can serve as guiding principles for a solution aiming to host and service AI models in a business application environment. These requirements are:

\begin{enumerate}
    % \item \textbf{User Experience} - Deployment of the AI models must be simple and easy, not requiring a high level of technical expertise.
    \item \textit{Target personas} - We define three target personas for KModels in the consuming end:
    \begin{enumerate}
        \item {\it Application administrators} - This persona is tasked with the basic installation of the framework and its initial setup. Later, this persona deals with any operational tasks such as configuring the hardware resources of the system and uploading new versions. The application administrators can also operate the model management however this task demands a certain level of business knowledge and it is likely that the application admin will delegate this task to expert business users (or carry it out on their behalf).
        \item {\it Expert business users / Subject matter experts} - The expert business users should provide the necessary business input and decisions for instantiating a new model.
        \item {\it Business users} - The end users of the business application. They will be viewing and utilizing the AI output in their daily routine. 
    \end{enumerate}
    We discuss additional personas involved in a typical KModel workflow later in this paper.
    \vspace{5px}
    \item \textit{Comprehensive lifecycle support} – The solution must provide integrated services covering the entire lifecycle of an AI model from the initial training, through data connectivity, monitoring, feedback, governance, versioning, caching etc.
    \item \textit{Scalability and extensibility} – The solution must support easy scaling of the system in the number of users/calls and number of models. Furthermore, the solution shall support easy extension of the system with additional services.
\end{enumerate}    

KModels is built upon the open sources frameworks: Kubeflow (or Data Science) Pipelines \cite{github2024kfpipe} and KServe \cite{github2024kserve} as core foundations ensuring scalability, efficiency, and industry proven standards (Figure \ref{fig:kmodels-stack}). As a result, many of KModels' design principles are based upon these frameworks.

\begin{figure}
\center\includegraphics[width=\linewidth]{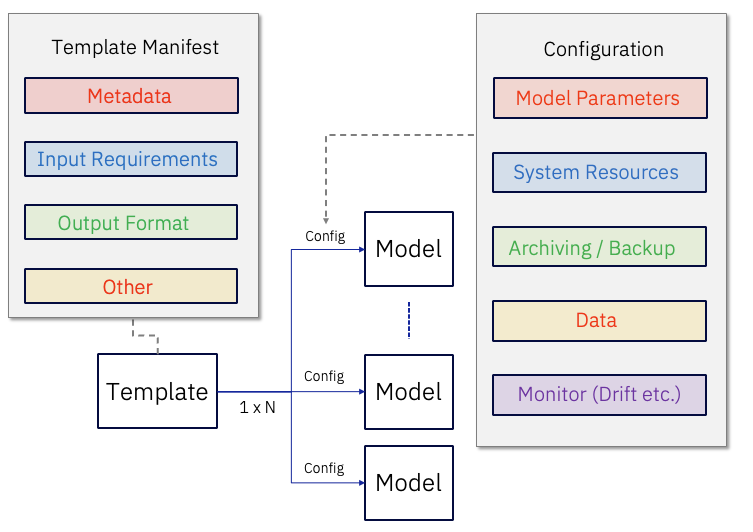}
\caption{The template's metadata (left) defines the information needed for instantiating a model and the supported arguments for configuration. Upon instantiation, a configuration file fills in all the necessary data and an model instance is created.}
\label{fig:config}
\end{figure}

\subsection{Model Templates}
\label{templates}

A cornerstone of the KModels solution is the concept of the model \textit{template}. The Template concept is analogous to the Docker concept \cite{poulton2023docker}, similar to how Docker containers can be spawned from a Docker image so can AI model instances be spawned from a model Template. A Template is an immutable package (file) encapsulating the source code, libraries, dependencies, tools, and other files needed for an AI model to operate. The creation of the template (essentially model development) and packaging is done on the model vendor's end. These steps require skilled personnel with data science and operational experience (Figure \ref{fig:kmodels-flow}). 

The template's code and metadata are developed by an AI developer (or data scientist) containing the necessary instructions for training and serving a model. The code is crafted in an iterative refinement process, and it is typically tailored to a specific business need, although this is not a requirement and AI developers may choose to define the boundaries of the use cases that they address. In fact, the KModels framework is completely agnostic to the internals of the template code and developers are free to write their code almost without any restrictions. 

Next, the model is packaged as a template image by an ML practitioner (it can also be the developer). Packaging the template is a one-time effort, per model, and it is also carried out on the vendor's end. Following the packaging of the template image, it can be transported to a client location, either directly or published in a central repository, enabling multiple clients to download and deploy it. Templates that have been deployed to the KModels framework can be used to create AI models. 

The next steps: Instantiation, training, deploying, etc., are all executed on the client's end. They are controlled and performed by the KModels framework and therefore require a relatively low level of skills, targeting personas of application administrators and expert business users, and not requiring any data science skills.

A formal and concise description of the way KModels and the Templates operate can be described in the following formulas:

% Version with MA(in_0^0,... C_0)
\begin{equation} \label{eq:Insantiate Template}
T(\tau, i, \bar{c}) \rightarrow \underset{\mathclap{\substack{\uparrow \\ C(\bar{in}, out, \bar{c}^0)}}}{\bigoplus}  \rightarrow UM^0(\bar{in}, out, \bar{c}^0, \tau, i)
\end{equation}
% Where \textit{T} is the template, \textit{\(\tau\)} is the training pipeline defined and implemented by the template developer, i is the inferencing pipeline defined and implemented by the template developer, and c is the available configurations that the template developer chose to expose to the client. 
% C is the configuration for instantiating the template into a specific model. It contains \textit{\(in_0,...,in_n\)} which specifies the input data for the model, \textit{out} which specifies the label to be learned in training and inferred from the input in run-time, and \textit{\(C_0\)} which is the specific setting for exposed configurations of the template. Configuration options that are not specified will receive the default values that the template developer defined.
% \textit{MA} is the model architecture (the untrained model). The untrained model is a specific instantiation of the template, prior to training phase. A single template could be used to create many different model architectures by use of different configurations.
% The Next step is the Training of the model. We divide this formally to the step of gathering and preparing the relevant data by the data preparation steps in the training pipeline, and then training the model architecture with the specific data to create a trained model:

% \begin{equation}
% T(\tau, i, c) \rightarrow \underset{\mathclap{\substack{\uparrow \\ C^0(c_0^0, ... c_n^0)}}}{\bigoplus}  \rightarrow UM^0()\}
% \end{equation}

\begin{figure}
\center\includegraphics[width=\linewidth]{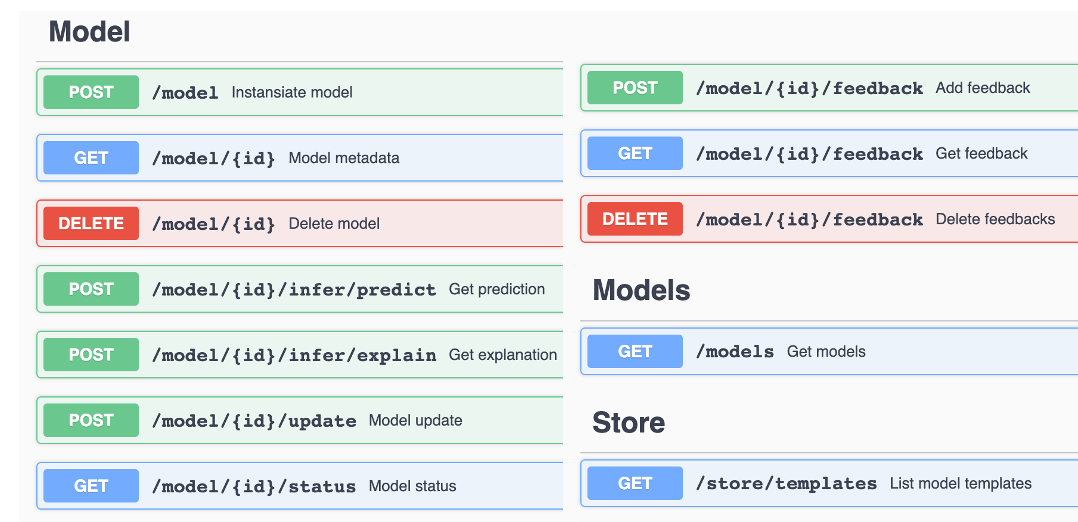}
\caption{REST interface for KModels allowing to create models, delete models, infer, provide feedback, etc.}
% This specific window shows the configuration for the Approval model.}
\label{fig:kmodels-rest}
\end{figure}

Where \textit{T} is the template, \textit{\(\tau\)} is the training pipeline defined and implemented by the template developers (i.e model developers in the context of template development), \textit{i} is the inferencing pipeline defined and implemented by the template developer, and \textit{$\bar{c}$} is the vector of available configurations that the template developer chose to expose to the client. 
\textit{C} is a specific configuration for instantiating the template into a specific untrained model. \textit{$\bar{in}$} is the definition of which data attributes (vector) the model should rely on for training and inferencing while \textit{out} defines the output of the model. For example, this could be names of specific columns or object attributes, depending on the product's schema. \textit{C} also defines \textit{\(c_0^0,...,c_n^0\)} which are the specific settings of the configurations to be used for the current model instance. Configuration options that are not specified will receive the default values that the template developer defined.
In all equations a superscript 0 indicates a specific instantiation.
% The output of this stage is the tuple \textit{\(\tau^0, i^0, MA^0\)} where \textit{\(\tau^0\)} and \textit{\(i^0\)} are the specific instatiations of the training pipeline and the inferencing pipeline created from setting \textit{\(C^0\)} to \textit{\(\tau\)} and \textit{i} respectively.  

\textit{\(UM^0\)} is the untrained model. The untrained model is a specific instantiation of the template, prior to its training phase. An untrained model contains the specific training and inferencing pipelines and their configurations, as well as the configuration of what data to apply the model on, the inputs and the output. A single template could be used to create many different untrained models.
\par
Next, the untrained model is trained on the subset of the client data that the untrained model is configured to operate on:

\begin{equation}
UM^0(\bar{in}, out, \bar{c}^0, \tau, i) \rightarrow \underset{\mathclap{\substack{\uparrow \\ D}}}{\bigoplus}  \rightarrow M^0(\bar{in}, out)
\end{equation}

Here, \textit{D} stands for the client data, and \textit{\(M^0\)} is the trained model, which has been trained on the data that was defined in the configuration, using the training pipeline defined in the template and the configuration captured in the untrained model.
The final step, which we formally define here mainly for completeness is the use of the model in run-time. This is a normal inferencing step:
\begin{equation}
\{M^0, D^0(\bar{d_0},..., \bar{d_m})\} \rightarrow \underset{Inferencing}{\bigotimes}  \rightarrow \{out_0,..., out_m\}
\end{equation}
Where the inferencing pipeline \textit{\(i^0\)} runs model \textit{\(M^0\)} on data points \textit{\(d_0,..., d_m\)} to generate the corresponding predictions \textit{\(out_0,..., out_m\)}. The inferencing could be done one data point and label at a time, the batch inferencing is shown as a general case.
\par

In order to standardize development process of templates and reduce the amount of unnecessary decision making for the template developer, KModels proposes a well-defined project structure. The template's project structure (Listing \ref{lst:template}) hosts two primary folders for storing the training code (\code{kfp}) and the serving code (\code{kserve}). The \code{template} folder stores the configuration metadata for the template. All of the following folders are optional and can be used only as needed. The \code{common} and \code{third\_party} folders store code libraries which are required both during training and during the serving of the model, however their affiliation is different as their names indicate. The \code{third\_party} folder is used to host external libraries while \code{common} stores proprietary libraries. All the aforementioned folders are folders which will be collected and packaged into the template images to be transported and consumed using KModels. The remaining folders will not be packaged and are mainly used by the template developer for local testing purposes during the template development phase. The \code{data} and \code{examples} folders can be used to store data samples and input samples and are both useful when running local tests of the template. The \code{hack} and \code{research} folders can store local scripts and code which is used as the template is being developed and different versions of the models are tested locally. The \code{pretrained} folder is used for storing the trained instance of the model when running local training. 

\begin{lstlisting}[
  backgroundcolor = \color{codegray},
  caption=The skeleton of a template project structure,
  label=lst:template,
  captionpos=b
]
|-- kfp
|-- kserve
|-- template
|-- common
|-- third_party
|-- data
|-- examples
|-- hack
|-- research
|-- pretrained
|- Makefile
\end{lstlisting}

In order to easily comply with this folder structure, KModels provides a boilerplate project, ready with the required folders and including bootstrap code suitable for creating a simple \textit{Hello, World!} style model. Additional to the folders, a template project comes with a \code{Makefile} in order to automate the process of packaging the template. The \code{Makefile} as well as additional \code{Dockerfile} are the "glue-code" that facilitates the construction of the template images in the correct order while collecting all the sub components. The persona which is tasked with customizing the Makefile (if needed), running it and finally producing the Template image is a skilled AI engineer or MLOps engineer. It is important to note that while the model development is typically carried out in iterative cycles (by a model developer), the packaging work of the AI engineer (also known as ML practitioner) is typically a one-off operation per a published version of the model's template.

A key point about templates is the fact that they are meant to be deployed, instantiated, trained and finally served without any intervention from a data scientist. Granted, a template will benefit from various runtime services, provided by KModels' infrastructure, that are required for its operation. However, it remains the duty of the template developer to instill within the template's code the robustness and flexibility that are needed in order to deal with a variety of data challenges (examples for that in section \ref{evaluation}). The template developers define the minimal input requirements for training the model. They can also expose to the business users, consuming the model, certain configuration parameters allowing to control business or even technical aspects of the model's operation (\textit{C} in Equation \ref{eq:Insantiate Template}). 

Beyond these configurations, the developers' responsibility is to write the code in a way that adapts well to data issues and tolerates potential issues such as poor data quality or changes in the data schema. Furthermore, the developers are expected to embed within the code some of the logic that a data scientist would have applied had they manually fine tuned the model. For example, the code should internally perform a train-test split and run multiple iterations of training, while altering the hyper-parameters, the input attributes or the size and split of the data. The code should evaluate many of these options. Eventually, it should expose the best suited model to be served by KModels. This requirement, as well as the packaging process of the model as a template, places a heavy burden on the shoulders of the vendors of the model. The time it takes to develop a model in a way that is generic and suitable to be served as a template is longer than the time it takes to develop a plain model. However, the potential benefit from being able to serve the model to multiple clients or multiple business applications is enormous and the value these clients may gain from being able to easily deploy the model without any coding is substantial.

\begin{figure}
\center\includegraphics[width=\linewidth]{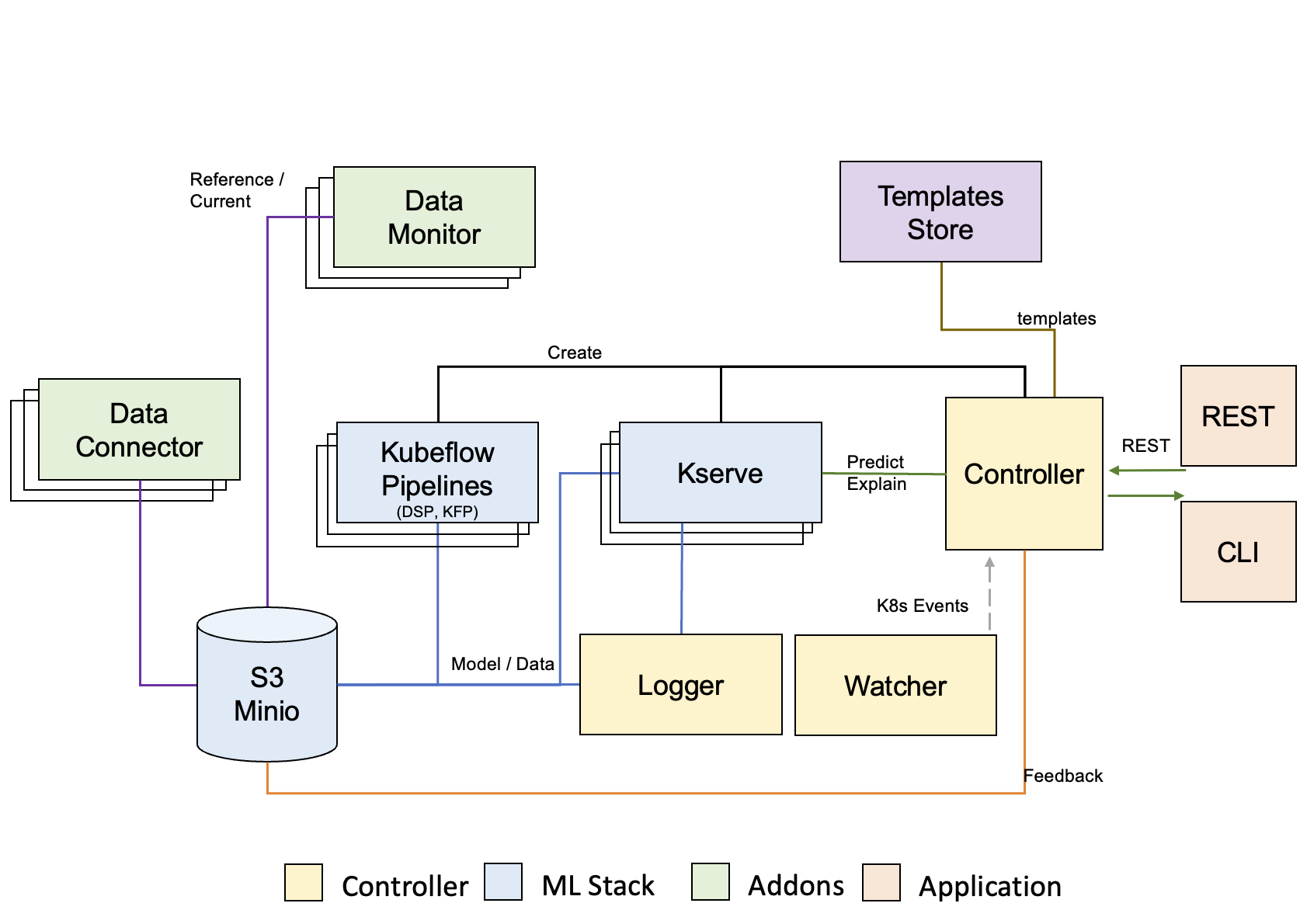}
\caption{KModels architecture}
% This specific window shows the configuration for the Approval model.}
\label{fig:kmodels-architecture}
\end{figure}

\subsection{Models Management}

After the template project is packaged into an image, it is transportable and can either be shipped directly to clients or be published in a \textit{Model Store}. The Model Store can reside in the same cluster with KModels, allowing for local consumption of templates or it can reside in a separate cloud deployment allowing multiple different KModels (clients) to connect to it, present the store's templates content and pull any of them locally, to be instantiated.

Once a template is within the KModels system, it can be used to generate multiple model instances (see Equation \ref{eq:Insantiate Template}). Each instance utilizes a unique configuration (Figure \ref{fig:config}). The business user will use the applications' GUI to set the different configuration options. 
% \underset{\mathclap{\substack{\uparrow \\ n = 2}}}{2}

The configuration defines:
\begin{enumerate}
    \item \textit{Resource parameters}: Resources consumed during training and serving. These allow the application admin (the persona tasked with creating the models on the client side) to allocate the necessary CPU, GPU and memory resources to each model. The template may define minimal requirements for launching a model. The admin has to ensure that the configured resources are indeed allocated to the KModels cluster.
    \item \textit{Data connection}: The specific connector and relevant data attributes. KModels comes out of the box with a set of predefined data connectors to a variety of databases and data sources (SQL, S3, file-system, etc.). The output format of the connectors is well defined and the template developers are aware of it when writing the data processing code. The data connectors allow the admins some level of control for filtering and selecting the input attributes form the data source. Mapping of the input data attributes to the template's specific input requirements is part of the template-specific arguments.
    \item \textit{Template-specific arguments}: These arguments are exposed by the template developers to allow control over different aspects of the model. Keeping in mind that the templates are aiming to hide the technical complexity from the users, while maintaining their control over the business behavior, we expect to see more business-centric arguments and less technical arguments. Furthermore, the more use-case specific a template is, the less arguments (flexibility) we expect to see.
\end{enumerate}

This configuration-driven flexibility allows application developers to tailor a template to a specific use cases with minimal user interaction. Having said that, it also support the provisioning of templates that are capable of addressing diverse needs. For example, imagine a template that encapsulates a generic capability of a binary classifier. This template should accept an input vector with multiple attributes of various data types and feed them internally to a binary classifier. It is likely that such a template would expose configuration arguments that support technical configuration of the model in order to better tune the model to the data. Such an open-ended template requires some level of ML understanding from the user.

Upon successful model creation, the KModels framework automates the entire lifecycle, encompassing: 
\begin{enumerate}
    \item \textit{Data Acquisition}: Fetching and managing necessary training data. KModels launches the respective data connector to pull the data and stores it for the model. It can also do so on a periodic basis for subsequent re-training of the model.
    \item \textit{Training Execution}: Leveraging Kubeflow Pipelines for robust and scalable training. KModels times the creation and launch of the training pipeline and monitors its execution.
    \item \textit{Model Deployment}: Efficiently serving the trained model using KServe infrastructure. Upon termination of the training pipeline, KModels will store the resulting model and configure KServe to serve the model. The application admins can intervene, based on the training metrics presented to them, and prevent the model from getting published to KServe.
\end{enumerate}

Beyond the essential lifecycle, KModels extends its support to comprehensive model management services: 
\begin{enumerate}
    \item \textit{Continuous Monitoring}: Tracking model performance and potential deviations.
    \item \textit{Drift Detection}: Identifying performance degradation or data drift. 
    \item \textit{Periodic and event based Retraining}: Automating scheduled model updates and Monitoring or Drift-Detection triggered retraining. 
    \item \textit{Archiving and Caching}: Optimizing storage and access for past models and intermediate results. 
\end{enumerate}

\subsection{KModels Architecture}
The entire operation of KModels is controlled by the \textit{Controller}. It serves as the "brain" of the framework, orchestrating the operation of the all other components. It triggers and monitors the operation of the components in a sequence, configuring each step and delivering the necessary data from one step to the next. Many of the components in KModels communicate using Kubernetes (K8s) events, hence the \textit{Watcher} listens on those and alerts the controller of the relevant events. 

The two primary components within KModels architecture (Figure \ref{fig:kmodels-architecture}) are \textit{Kubeflow Pipelines} \cite{github2024kfpipe} (or Data Science Pipelines) handling the entire training flow and \textit{KServe} \cite{github2024kserve}, serving the models. These components are widely accepted in the industry as a solid infrastructure for AI deployments. However, in a typical AI deployment, these components require highly skilled MLOps engineers in order to set them up and configure the AI models to run on them. Using KModels, their installation, setup and configuration is handled by the \textit{Controller}, significantly minimizing the overhead involved in their operation. 

KModels uses an S3-compliant storage for storing the data, models, and any other metadata required. The S3 standard defines a set of application programming interfaces (API) for storing data over a cloud infrastructure and is largely based on Amazon's S3 architecture \cite{s32024amazon}. KModels may be installed out of the box with a local S3 storage - Minio \cite{github2024minio}. Alternatively, one can provision any S3-based service and provide its credentials to KModels during installation. 

The \textit{Data Connectors} and the \textit{Data Monitors} are two sets of pluggable components supporting the extension of KModels. KModels provides an internal API for these services allowing them to receive notifications about key events in the lifecycle of the model and to gain visibility into the data. 

\begin{figure}
\center\includegraphics[width=\linewidth]{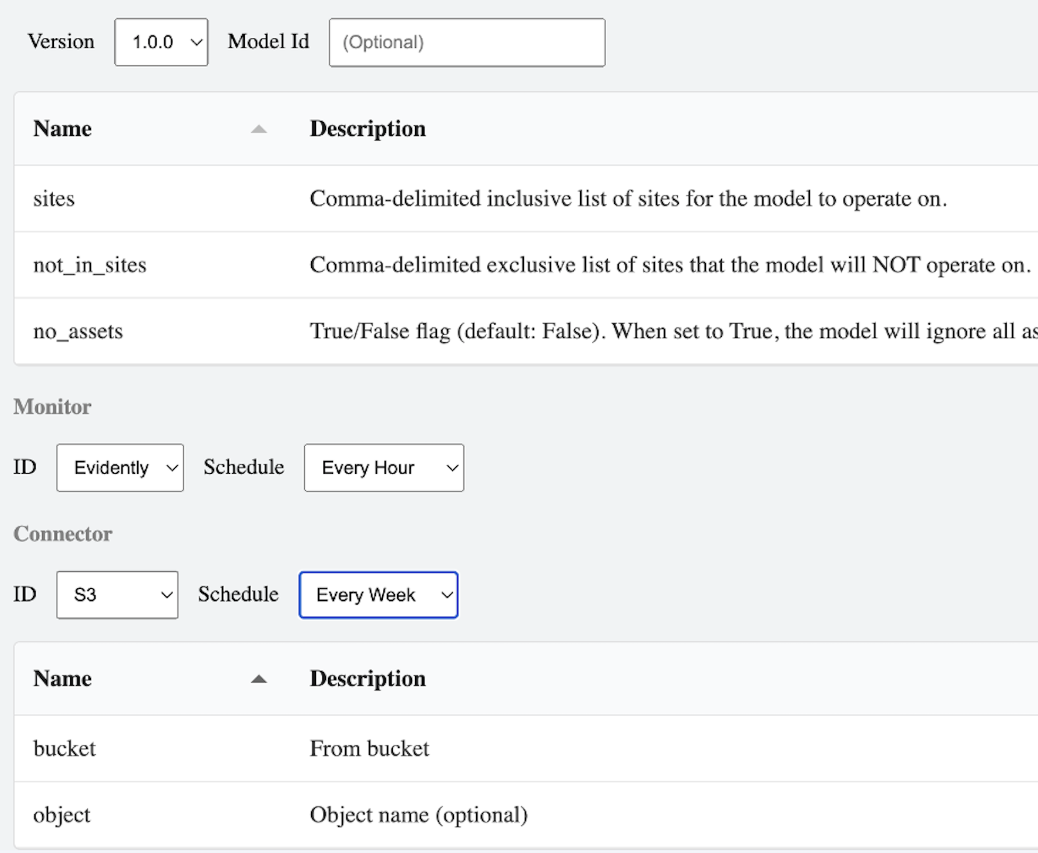}
\caption{Applicative GUI developed for a client pilot setting, allowing the users to manage the AI models in a user-friendly manner. This example is for an Approval model.}
% This specific window shows the configuration for the Approval model.}
\label{fig:dashboard}
\end{figure}

The data connectors are tasked with fetching the data from any external source and serving it to the model at training time. Once a data connector is configured for a model instance, it will receive the necessary configuration from KModels and be triggered to operate at the relevant point. The data connectors are familiar with the logic of the data source they are targeting: How to connect to it, its schema, attributes, etc. They are responsible for mediating the source data format to the desired input format of the models. In order to promote alignment KModels defines a standard data format so that both template developers and connectors developers can use it. Having said that, this standard is merely a reference and specific template or connector may define their own suitable format for data. 

The data monitors are expected to track the internal distribution of the data (and the predictions) over time and alert for any deviations. The monitors will be given access to the training and inference data and will apply internal logic to analyze the data for any potential issues. As an example, the drift detection service is an extension of KModels based on an open-source library, Evidently \cite{github2024evidently}.

KModels exposes a control API allowing application administrators to control template and model management. The API is exposed in the form of a REST interface and a command-line interface (CLI) (Figure \ref{fig:kmodels-rest}). The applications are advised to either wrap this API in a graphical user interface (GUI) or alternatively to automate the model management calls, while hiding the control logic from the end-users. Contrary to that, that applications \textit{must} provide a GUI for the business users reviewing the models outputs. The reasoning is that, unlike the application administrators, that business users should not be required to use REST or the CLI to interact with the AI. All their interaction is GUI based and embedded in the application UI, while using the application terminology.

% Eyal - need to enrich the Deployment chapter and review the architecture chapter

\section{Deployment}
KModels is deployed on a Kubernetes or OpenShift clusters, typically on-premise on a client's local data center but possibly also on a cloud data center. The installation requires a foundational layer of an AI platform. On OpenShift that layer is \textit{Open Data Hub} while on Kubernetes it is represented as \textit{Kubeflow} (Figure \ref{fig:kmodels-stack}). Next, KModels is installed, either using manifest scripts \cite{github2024kubeflowmanifest} or an OpenShift Operator \cite{github2024odhoperator}. In practice, KModels only utilizes two services from the AI platform - Kubeflow (or Data Science) Pipelines, for the training pipelines and KServe for serving the models. Overall installation time for KModels and the AI platform typically takes less than 30 minutes.

KModels framework can run on a single-node cluster with minimal resources. The resources that the framework consumes can conceptually be split to three types: AI platform resources, KModels resources, and models runtime resources. The first two types are mostly static and invariable to the number of models. In a minimal deployment, the system utilized as little as 4 CPUs and 6GB of memory, without any running models. As the system ramps up model instances, so do the resource requirements ramp up as well. These runtime requirements depend heavily on the type of models and their specific memory and CPU requirements. For example, a simple XGBoost classifier for tabular data may require less than 1 CPU and 1GB while on the other end a \textit{Large Language Model} (LLM) may require a significant amount of CPUs and memory as well as GPUs. KModels allows the application administrators to limit the resource allocation on a per-model basis using the configuration given during model instantiation. Furthermore, using KServe for inferencing implies that the system automatically scales its runtime allocations pending on the actual requirements of the running models. KServe enables auto-scaling based on request volume and supports scale down to and up from zero.

\begin{figure}
    \centering
    \includegraphics[width=\linewidth]{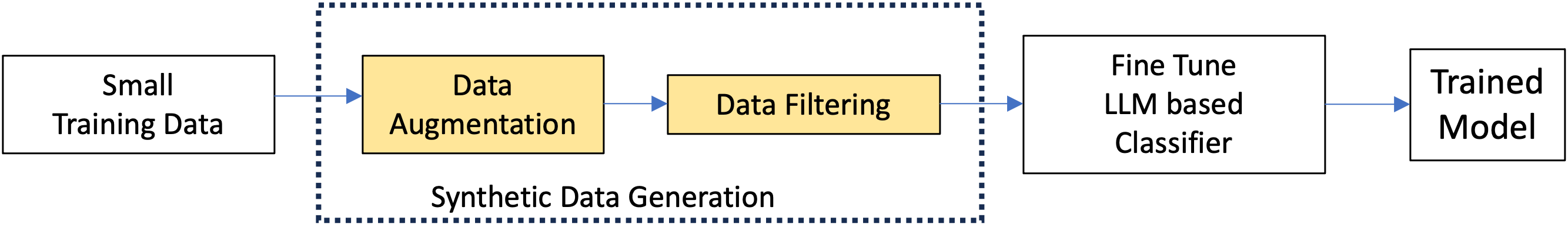}
    %(b)\includegraphics[width=0.7\linewidth]{Figures/generic_example.png}
    \caption{Example of a failure code recommendation model generation pipeline.}
    % and (b) synthetic data generation.}
    \label{fig:fcc}
\end{figure}
\begin{figure}
    \centering
    \includegraphics[width=\linewidth]{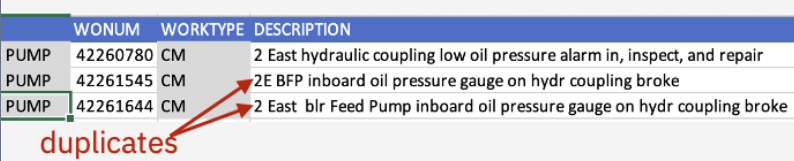}
    \caption{Similar work order suggestions using temporal and textual information.}
    \label{fig:similarity}
\end{figure}

After KModels is deployed, the application administrator can browse the KModels template store, inspect the templates' metadata and select the relevant templates for model deployment. Model creation involves customizing configurations (Figure \ref{fig:config}), mapping local data to the template's input fields and providing additional arguments as business context (Figure \ref{fig:dashboard}). Users are able to experiment with multiple different configurations, easily creating multiple model instances from any single template. This flexibility allows the clients to adapt models on-site to diverse needs, such as applying different filters and temporal windows to incoming data. Additionally, numerous (dozens and more) models can be easily spawn with different configurations based on the same template, allowing the creation of models per specific data pattern. For example, one may train a separate model for different asset classes.
The time it takes to configure and launch a new model is typically less than a few minutes. The training time and the time up till the model is ready for inferencing, depends on the model's template and the allocated system resources. Once the model training is launched, no user interaction is required.

%......

\section{Evaluation}
\label{evaluation}
KModels was used to deploy three different AI models in an existing asset management application for a global real estate business. These model templates use three completely different underlying approaches with varying complexity and AI methods. The three models are:
\begin{enumerate}
    \item Failure Code Recommendation (FCR): A multi-class classification model (including synthetic data generation); Recommends the most suitable code (from a total of usually 30-60 codes) based on the description (Figure \ref{fig:fcc}) text within the records.
    \item Similarity Suggestions: This model identifies similar work orders (Figure \ref{fig:similarity}) based on a combination of categorical, spatial, temporal, numerical and textual features. Cosine similarity of contextual embedding is used for textual data processing.
    \item Approval Decision Recommendation: A binary classification model generating a recommendation to approve or reject a record \cite{abitbol2023improving}. This model utilizes a wide feature vector based on the records tabular data.
\end{enumerate}

% TODO: Add more information about the different techniques used to create these models. This should then be used to emphasize the flexibility of the framework, that enables the data-scientist to use different techniques and doesn't limit to a specific type of model.

% TODO: Bhavna - please add here aparagraph about the key principles of the model

% As shown in Figure \ref{fig:fcc}, the FCR model generation involves the option to create synthetic data in case there is insufficient data to fine tune a multi-class classifier based on encoder only language models \cite{devlin2018bert}. Fine-tuning the classifier was important because this application's data is very domain-specific (sometimes even client specific), and thus mostly unsuitable for zero-shot LLM-based classifiers like \cite{meng2022supergen}. The underlying synthetic data generation is guard railed based on the available data, and uses multiple approaches that invoke other generative and non-generative LLMs, like round trip translation \cite{sennrich2015rtt}, masked language modeling \cite{kobayashi2018mlmlike} and generative approaches using prompt engineering. This generated data is then filtered using sentence transformers \cite{reimers2019sentence} and COVE-like \cite{dhuliawala2023cove} approaches. And finally the standard encoder-only language model is fine-tuned for the multi-class classification task, generating a final classification output.

The FCR template has a complex architecture (Figure \ref{fig:fcc}) consisting of multiple steps of varying complexity, including several data augmentation techniques (\cite{devlin2018bert}, \cite{meng2022supergen}, \cite{sennrich2015rtt}, \cite{kobayashi2018mlmlike}) with touch-points to multiple LLM models. However, this complexity is hidden from the end users allowing them to provide merely a minimal set of configurations so that KModels runs the complex pipeline on their behalf.

% The Similarity suggestions model compares work orders using categorical, spatial, temporal, numerical and textual features. It collects and concatenates text from a number of text fields in the work order record and creates sentence embedding using Sentence Transformers \cite{reimers2019sentence}. In order to identify potentially (textually) similar records it applies several similarity techniques on the embedding vector (e.g. cosine similarity). 

The Similarity template's configuration supports different use cases using control and terminology which are familiar to a business user. For example, identifying recurring issues in a given time-window can be configured as: "compare to recently closed records" within a certain "time-window size". This can further trigger investigation to identify root causes. In another case, alerting for duplicate records that are opened for the same problem, in order to prevent sending two technicians to address the same problem can be done with a setting of: "compare to open tickets". A different setting using "compare to completed tickets" may support the case to retrieve historic records for similar issues, to provide information on how these issues have been addressed in the past. 

% The Approval template is based on an XGBoost \cite{chen2016xgboost} classifier. The pipeline for this model is capable of ingesting a feature vector of varying length with different data types. 

% Internally, this pipeline performs several pre-processing steps, including: Standardizing the data, dealing with anomalous data points and missing data, transforming categorical data into numeric representation (using CatBoost \cite{prokhorenkova2018catboost}), feature engineering of date attributes, etc. The approval template performs an internal evaluation of the quality of the resulting model by splitting the data to train and test sets, training only on the train and assessing the performance of the model on the test data. 

% This Approval template internalizes several policies in order to improve the resulting model: Data standardization, imputation, cleanup, feature engineering and encoding (CatBoost \cite{prokhorenkova2018catboost}). This template performs multiple training iterations, each with a different set of input attributes and different train-test splits. For example, time-sensitive data is split multiple times, each with a different duration, in order to test the best training duration corresponding to the latest test period. In essence, the typical practices for a data scientist operating in the lab are embedded into the template code, effectively reducing the required skill level of the end user. 

The approval template exposes a configuration that allows the business users to tailor the model to their selected cohort. A default setup will not apply any filtering on the template allowing the instantiated model to run on the entire range of the data, producing a "global" model. An alternative setup, using a filter on the incoming data allows the model to train on specific sites, thereby supporting a notion of locality and allowing the creation of multiple "small" models each catering better to the data derived from its own locality. Users may leverage this ease of configuration in order to spawn numerous models, with little effort.

These diverse use cases demonstrate the versatility of the KModels framework, the flexibility it offers the template developer to work with different types of models and pipelines, and the simplicity it allows the end users to operate and consume the models. The ability to easily alter the configurations and produce multiple models supporting different use cases further underscores this flexibility and the empowerment that KModels offers to its business users.

The model code underwent rigorous development and testing using multiple client datasets to ensure robust performance. Through several iterations of offline testing, the processing pipelines were refined to handle data inputs of varying quality levels. This iterative process significantly improved the models' ability to adapt to diverse real-world scenarios.

Each model was then transformed into a KModels template by the development team. The conversion process varied in complexity and duration across the different models. The Approval model, being relatively straightforward, took only three days to convert. In contrast, the more complex Failure Code Recommendation (FCR) model required up to two weeks of effort to fully adapt it as a KModels template. This range in conversion times reflects the varying complexities of the models and the care taken to ensure they would function effectively within the KModels framework. However, in all of these cases, we consider this one-time effort as a minor investment compared to the simplicity gained in deploying them on multiple sites. The ability to easily replicate and customize these models across various client environments significantly outweighs the initial conversion effort.

% \begin{figure}
% \center\includegraphics[width=\linewidth]{Figures/workorder.png}
% \caption{The Work Order management process features several potential decision points suitable to be supported by an AI model. We selected three of those points and implemented them as AI model templates.}
% \label{fig:workorder}
% \end{figure}

\begin{figure}
    \centering
    \includegraphics[width=0.7\linewidth]{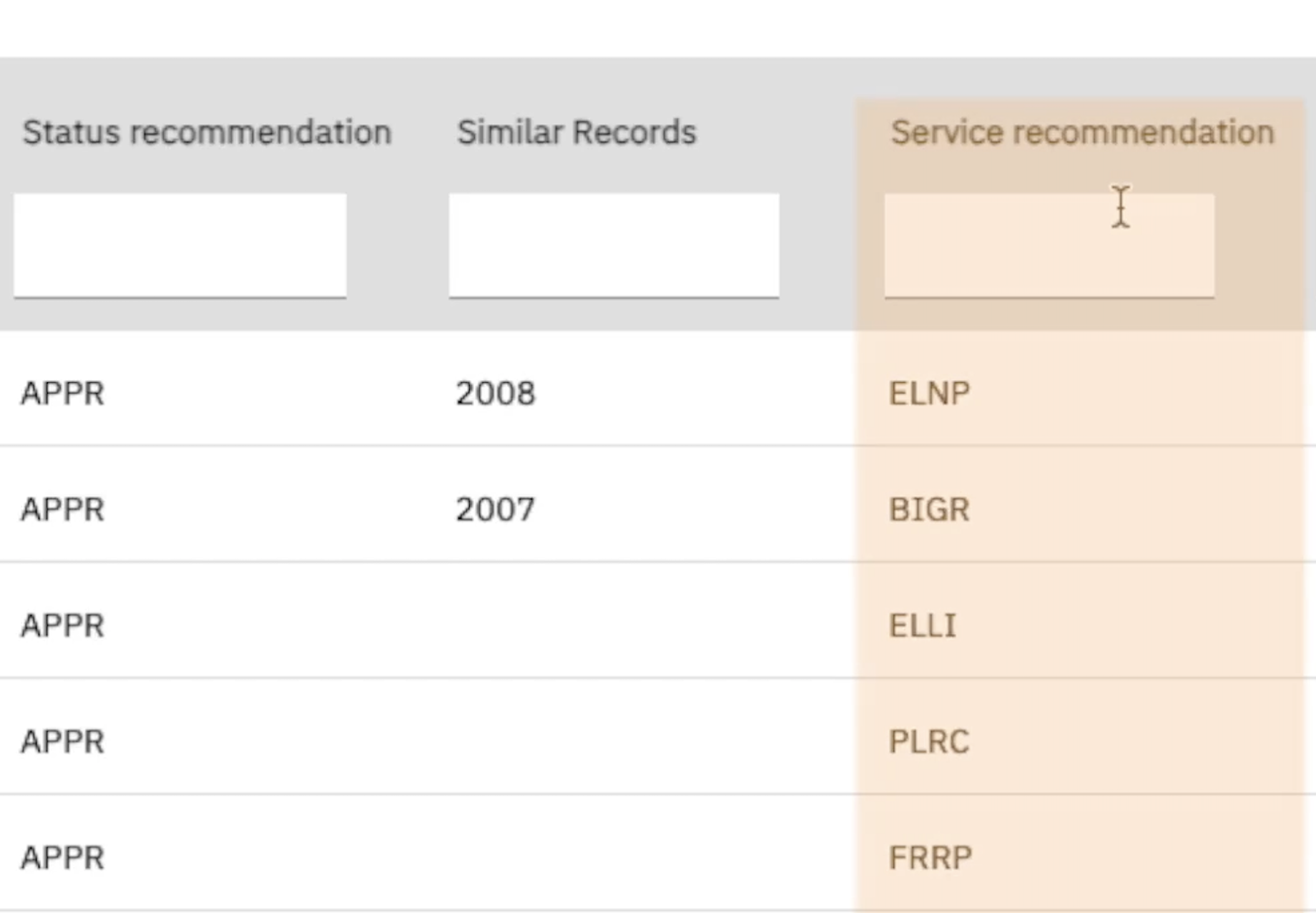}
    \caption{Consuming the recommendations of the AI models: Approval (Status recommendation), Failure Code (Service recommendation), and Similarity in the application.}
    \label{fig:output}
\end{figure}

The models were deployed using KModels along side an existing asset management application, at a client's site, with a throughput of several hundred newly opened work orders per day. The client made minor changes to their GUI to surface the recommendations to the end-users, as reflected in the table columns in Figure \ref{fig:output}. The initial setup and installation of KModels at this client's site, took roughly one hour to complete. Then, the application administrator of the application used the provided CLI in order to instantiate and launch the training of the three models, within a few minutes (See demo in the supplementary material). The administrator acknowledged that this technique is a huge leap forward in terms of the overall time they invested when deploying other AI services in the past. 

% The models were deployed using KModels alongside an existing asset management application at a client's site that processes several hundred newly opened work orders daily. To integrate the AI recommendations into the workflow, the client made minor modifications to their graphical user interface, as illustrated in Figure \ref{fig:output}. The initial setup and installation of KModels at this client's site was quite easy, taking approximately one hour to complete.

Following the installation, the application administrator utilized the CLI to instantiate and initiate the training of all three models. This process was completed within minutes, demonstrating the streamlined nature of KModels (a demonstration video is available in the supplementary material). The administrator emphasized that this approach represents a significant advancement in terms of time investment compared to their previous experiences deploying AI services. This substantial reduction in deployment time and complexity underscores the practical benefits of the KModels framework in real-world business environments.

While this current deployment has successfully handled hundreds of work orders per day, KModels' architecture is designed to scale to much larger workloads, potentially supporting thousands of models across multiple business units in large enterprises.

Following the introduction of the models' recommendation, the Failure Code specification for work orders improved from 46\% to 83\%, showcasing the substantial benefit of accessible and localized AI solutions. While this improvement doesn't directly demonstrate the capabilities of KModels, it illustrates that with the framework provided by KModels, a model deployed "into the wild" without prior training can autonomously learn from local data and achieve good performance, even with minimal oversight from an application administrator.

% All three model templates were initially developed in our lab, in a controlled setting. We received a large dataset, reflecting more than a year's data, from our partnering client and we set out to develop the best model for each of the tasks, given the business requirements. We conducted several iterations of train, validation and testing on a fresh set of client data and as the models matured and their results were satisfactory and stable we followed the KModels template guidelines and packaged the models into templates, allowing them to be transported and consumed on a target KModels framework.

% Once the client's site was fitted with a KModels installation, we uploaded the templates and allowed users to use the system. Deploying models on the client site was effortlessly enabled by the KModels framework. 

\section{Limitations and Threats to Validity}
While KModels offers significant advantages for AI adoption in business applications, it is important to acknowledge potential threats to validity and limitations of the approach:
\begin{enumerate}
    \item Template Development Overhead: The primary challenge lies in the additional effort required from model developers to align their work with KModels' template structure and requirements. This overhead may initially slow down the development process, though it potentially leads to more robust and reusable models in the long term.

    \item Installation Complexity: Deploying KModels introduces an additional component to the business application's infrastructure. This may increase the complexity of the initial setup and maintenance processes. However, this added complexity is offset by the simplified AI model management that KModels provides.

    \item User Interface Development: Application developers need to create a user interface to interact with KModels' APIs. While this requires some additional development effort, it allows for customized integration that aligns with the consuming application's look and feel.
\end{enumerate}
It is worth noting that these challenges are not unique to KModels. In fact, they often present even greater obstacles when implementing AI capabilities without a standardized framework. KModels aims to mitigate these issues by providing a structured approach to AI model deployment and management.

\section{Summary}
KModels reduces the barriers hindering widespread AI adoption in business applications. First, model developers that package their models as templates can easily source them to multiple clients and applications. Second, application developers can easily infuse their business applications with a variety of AI models without having to deal with the complexity of operational requirements and services. And last, clients benefit from models trained specifically for on their local data, on-premises, without requiring data science skills.

Finally, KModels' unique combination of capabilities is unmatched by other AI deployment frameworks: 
\begin{enumerate}
    \item Data remains on the local environment. On-premises model training without sharing data and with no data science skills.
    \item A template-based approach simplifying deployment and potentially accelerating AI adoption.
    \item Business-centric configurations for non-technical users to deploy and fine-tune models.
    \item On-site adaptation via customizable configurations.
    \item Foundations in Kubeflow Pipelines and KServe ensuring scalability, efficiency and standardized deployment.
    \item Comprehensive lifecycle management services.
\end{enumerate}

% KModels framework manages all resources related to the models. It will allocate storage to the model for storing the input data and the model's binary output and it will allocate processing power (CPU or GPU) for training and for serving the models. Naturally, KModel's resources are limited to what it was allocated with, during the initial installation of the framework. Each new model instance will receive, by default, a small allocation of storage and processing units, however, as part of the configuration, users can override that and request for larger allocations. Upon exhaustion of the KModels system resources it will no longer allow the creation of new models.
\bibliographystyle{IEEEtran}
\bibliography{kmodels}

\end{document}